# Spontaneous emission of electric and magnetic dipoles in the vicinity of thin and thick metal

R. Hussain[1], D. Keene[2], N. Noginova[1,*], M. Durach[2,**]

[1] *Norfolk State University, Norfolk, VA 23504*
[2] *Georgia Southern University, Statesboro, GA 30458*
*nnoginova@nsu.edu, **mdurach@georgiasouthern.edu

**Abstract:** Strong modification of spontaneous emission of $Eu^{3+}$ ions placed in close vicinity to thin and thick gold and silver films was clearly demonstrated in a microscope setup separately for electric and magnetic dipole transitions. We have shown that the magnetic transition was very sensitive to the thickness of the gold substrate and behaved distinctly different from the electric transition. The observations were described theoretically based on the dyadic Green's function approach for layered media and explained through modified image models for the near and far-field emissions. We established that there exists a "near-field event horizon", which demarcates the distance from the metal at which the dipole emission is taken up exclusively in the near field.

## 1. Introduction

The effects of the local environment on spontaneous emission are commonly discussed in terms of the Purcell effect [1] accounting for a modification of the photonic mode density and a subsequent alteration of the dipole emission rate [2-5]. Depending on the degree of modification of electric and magnetic components of optical modes, electric and magnetic dipoles can be affected in a different manner. This was discussed theoretically [4-8] and shown experimentally by observing changes in luminescence spectra of rare earth ions such as $Eu^{3+}$ [9-18], having both magnetic and electric dipole transitions. It was suggested that $Eu^{3+}$ ions can be used as a spectroscopic tool for probing the effect of optical magnetic resonance in plasmonic nanostructures [19], and for mapping local distributions of optical magnetic and electric fields in plasmonic metamaterials [15]. It was established that losses in nanostructured materials and changes in radiation patterns, which are different for magnetic and electric dipoles, are important factors for these applications [10-12].

Modification of electric and magnetic dipole emission associated with the presence of metal is an open problem in nano-optics and has recently attracted a lot of attention [12, 13]. If an emitter is placed in the vicinity of an ideal mirror and oriented parallel to the interface, one can expect a reduction of an electric and an enhancement of a magnetic dipole emission normal to the interface due to the boundary conditions for optical electric and magnetic fields [8, 17]. However, in very close vicinity to real metals at distances of about 30 nm, the opposite behavior has been recently observed: the emission of the electric dipole was enhanced while magnetic dipole emission was decreased near thin gold films and nano-strip arrays [10, 18].

The goal of the current work is to provide a better understanding of the effects of close vicinity of metal on electric and magnetic emitters. Here we restrict ourselves to planar geometry, considering dipoles very close to the surface of thin and thick metal films. The paper is organized as follows. First, we describe an experiment where the distinctly different behavior of electric and magnetic emitters located near thin gold films was visualized in an optical microscope setup. Then, we provide a theoretical description where we show that that the contribution of $Eu^{3+}$ emitters to far-field radiation demonstrates a threshold-like behavior dependent upon the distance between the emitters and the metal surface. In very close vicinity to the metal, all of the energy imparted on the emitter is required to establish a near field image within the metal, leaving nothing for radiation into the far field, which we refer to as being beyond the "near-field event horizon". Our model establishes a theoretical framework for the estimation of this threshold as a function of the thickness of the metal film. Also we show that it provides an adequate description of the effects observed in far field emission, which was originated from emitters located outside of this "event horizon."

## 2. Experiment

Highly luminescent $Eu(TTA)_3(L18)$ chromophore material was synthesized in house, following Ref. [20]. The emission spectrum of $Eu^{3+}$ has several well-distinguishable spectral lines, Fig. 1. The transition $^5D_0$ - $^7F_1$ with the emission at the wavelength, $\lambda = 590$ nm is associated primarily with a magnetic dipole [20] while the rest of the lines are primarily electric dipole transitions, including the strongest line, $^5D_0$ - $^7F_2$ with $\lambda = 611$ nm, originating at the same energy level.

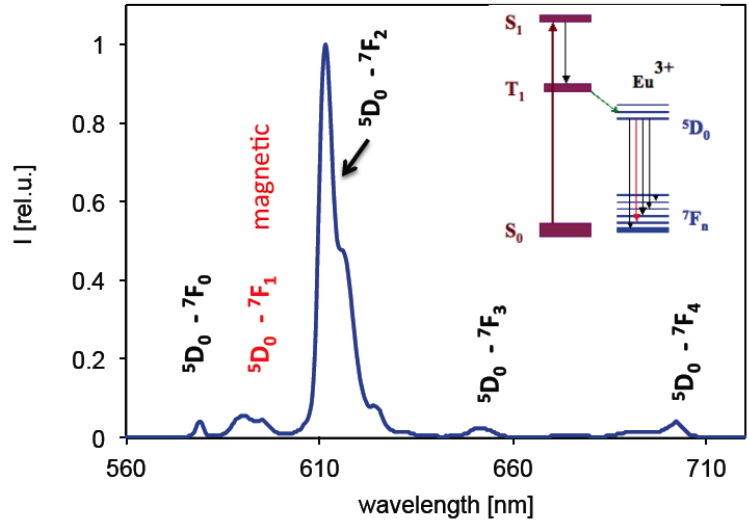

Fig. 1. Emission spectrum of $Eu(TTA)_3(L18)$ amphiphilic complex. The excitation wavelength is 330 nm. Schematic of the levels is shown in insert.

The idea behind our experiment was to use the microscope setup where one could simultaneously observe the emission of $Eu^{3+}$ placed in different surroundings: near thin metal, thick metal, and glass, which would be used as a reference. Then we would record and compare the effects of the different placement on the emission intensity separately for magnetic and electric transitions.

The substrates were fabricated with thermal deposition of gold or silver on a glass substrate through a standard STM mesh, 656-300-AU, purchased from *Ted Pella Inc*. Such a deposition produced 7 μm x 7 μm square patches of metal with 2 μm distances between each other, arranged in square blocks of ~50 x 50 μm size with 15 μm distances between blocks. The thickness of metal after the first step of deposition was ~ 50 nm as measured with the *Bruker DektakTX* profilometer. In order to obtain metal squares of two different thicknesses on the same substrate, we covered a half of the sample, and continued the thermal deposition. After the second phase, the thickness of squares at the exposed part was in the order of 170 nm.

Solutions of $Eu(TTA)_3(L18)$ complex and polystyrene in chloroform were mixed in the proportion 1:5. 30-microliter drop of the mixture solution was spread on a water surface. After evaporation of chloroform, a thin polymeric film was formed on the water surface. Such a process produced films with practically uniform thickness (which was confirmed with the profilometer after transferring the film to a flat surface). Immersing the substrate with metal squares, the film was transferred to the substrate covering both squares and a space between them. The thickness of the $Eu^{3+}$ polymeric films was in the range of 30-40 nm.

The microscope images were recorded using *Zeiss Imager Z2m* microscope equipped with *Axiocam* camera. The luminescence of $Eu^{3+}$ was excited with UV light at $\lambda = 325$ nm, which was brought to the sample with the optical fiber from the CW He-Cd laser. In order to record the emission signals at electric and magnetic transitions separately, interferometric filters for 610 nm and 590 nm correspondingly were inserted in the recording channel. The signal at 590 nm was relatively weak, that restricted us to use the 20x resolution objective of microscope.

The images obtained in the sample with thin gold are shown in Fig. 2. In Fig. 2 a, the golden squares seen in the standard reflection mode (using the microscope light source) correspond to square arrangements of small gold patches. The total emission, Fig. 2b, is brighter on the gold than on glass between them. However, the image clearly shows the presence of the luminescent film on both gold and glass.

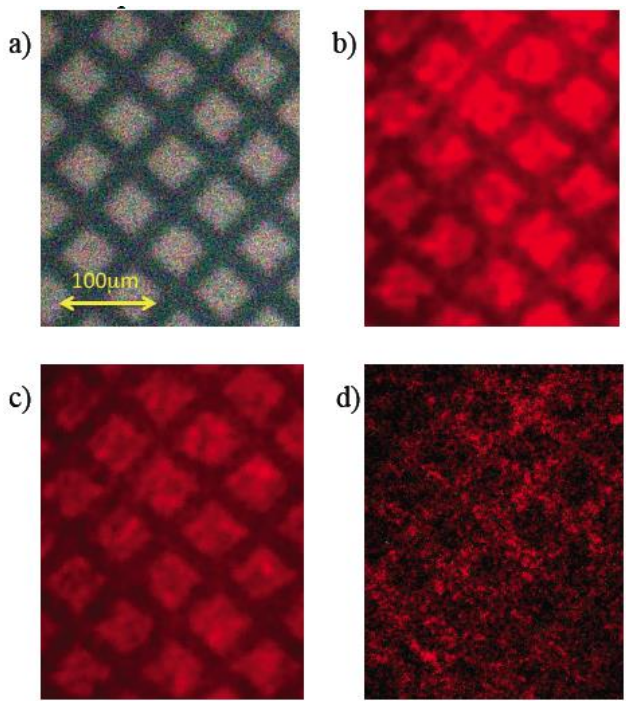

Fig. 2. a) A substrate with thin gold squares in a standard reflection mode; $Eu^{3+}$ luminescence: b) total; c) at 610 nm; d) at 590 nm

Images taken at 610 nm (strong electric transition) and 590 nm (magnetic transition) are shown in Fig. 3(c) and Fig. 3(d) correspondingly. As one can see, the image recorded at the electric dipole transition (Fig. 3(c)) is similar to the image with the total emission (Fig. 3(b)), which can be expected taking into account that the transition at 610 nm contributes of ~ 70% to the total signal. For the magnetic dipole emission, the contrast between gold and glass is the opposite (Fig. 3(d)): the film on gold is darker than on glass interspacing.

Such a difference in contrasts for magnetic and electric dipole emission exists only if gold is thin (50 nm). At larger thicknesses of metal, both electric and magnetic dipoles show similar behavior. In order to clear demonstrate this, the polymeric film with $Eu^{3+}$ was deposited onto a substrate having both thick and thin metal patches in such a way that the polymeric film of almost uniform thickness covered both thick and thin metal patches and bare glass.

In Fig. 3a, recorded in the standard reflection mode, different thicknesses of gold squares can be distinguished by different colors of squares. The light colored squares (indicated with a circle at the top of the figure) were thicker (d ≈170 nm) and dark colored squares (bottom circle) were thinner (d ≈50 nm), Fig. 3(a). The images recorded at 610 nm and 590 nm are shown in Fig. 4((b), (c)). Some variation in the emission from top to bottom is related to non-uniform illumination due to the position of the excitation source. However, the character of contrast between gold squares and glass in the inter-space is clearly seen.

The image recorded at the electric dipole transition (Fig. 3(b)) shows much stronger emission intensity from the $Eu^{3+}$ placed on the top of gold squares than that on the glass (inter-square spacing). The character of contrast does not depend on the thickness of gold: gold brighter than glass is seen for both thin and thick patches.

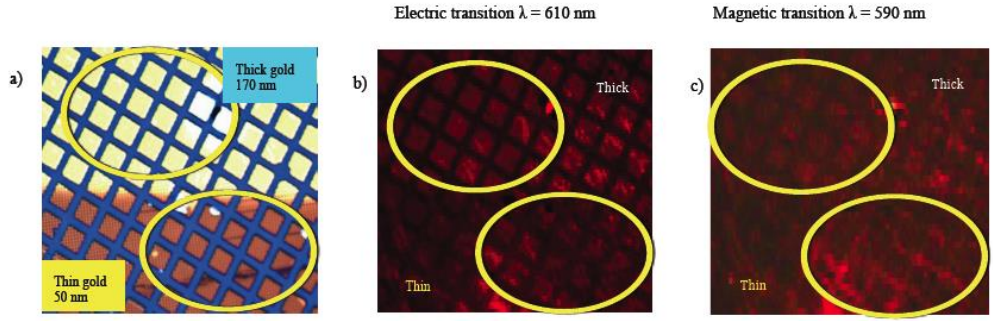

Fig. 3 a) Thick and thin (as indicated with circles) patches of gold on glass in reflected light. b) and c) $Eu^{3+}$ luminescence at 610 nm and 590 nm correspondingly

The magnetic transition (Fig.3(c)) shows the negative contrast (gold is darker than glass) only for the thin gold (see the bottom circle). The contrast between thick gold and glass was similar to what was observed for the electric transition (gold is brighter than glass, see squares in the top circle).

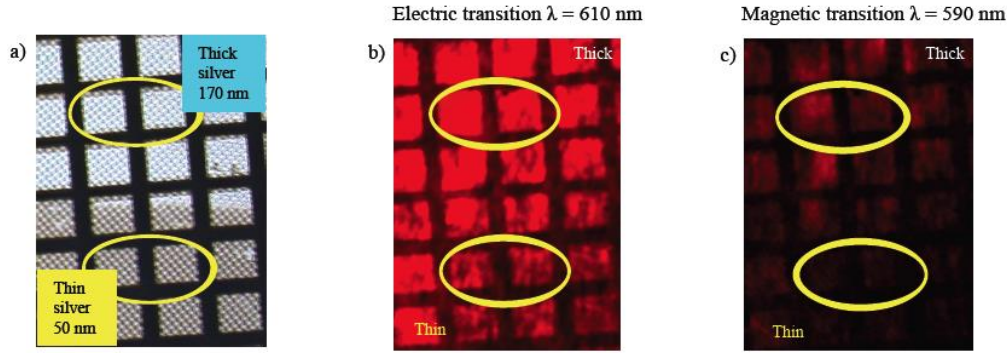

Fig. 4. a) Thick and thin (as indicated with circles) patches of silver on glass in reflected light. b) and c) $Eu^{3+}$ luminescence at 610 nm and 590 nm correspondingly

We repeated the same experiment using a similar substrate having thin and thick silver patches, Fig. 4. In opposite to the observations with gold, the contrast was the same in all cases. For both electric and magnetic transitions, thick and thin silver squares looked brighter than glass, however, the magnetic dipole emission was significantly weaker on the top of thin silver than that on thick silver.

## 3. Theory

Our formulation is based on the dyadic Green's function approach for layered media [22]. Consider the structure composed of a glass substrate with refraction index $n_g$, a metal film with thickness $a$ and a polymer layer with thickness $d$ and refraction index $n_p$, containing a dipole separated by distance $h$ from the metal (see Fig.5). We show that the behavior of the emitters is strikingly different depending on the parameter $h$.

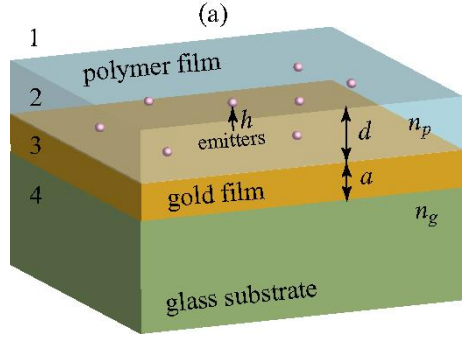

Fig. 5 Schematic of the structure.

The experiments are performed at CW UV excitation, which implies a steady state condition $P_{uv} = P_e + P_m + P_{non-em}$ where the excitation power $P_{uv}$ is equal to the power released by the ions in the form of electric dipole emission $P_e$ at the transition $^5D_0$ - $^7F_2$, magnetic dipole emission $P_m$ at the transition $^5D_0$ - $^7F_1$ as well as $P_{non-em}$ released through other radiative and non-radiative channels.

In this paper we use normalized emission rates $F_e$ and $F_m$ defined through $P_e = \hbar\omega_e\Gamma_{0e}F_e$ and $P_m = \hbar\omega_m\Gamma_{0m}F_m$. Here the spontaneous emission rate for electric and magnetic dipoles in a homogeneous polymer medium are $\Gamma_{0e} = 4k_0^3|d_e|^2/(3\hbar n_p^2)$ and $\Gamma_{0m} = 4k_0^3|d_m|^2/(3\hbar)$. The normalized emission rates are equal to the integrals $F_e = \int_0^\infty \tilde{\rho}_e(k)dk$ and $F_m = \int_0^\infty \tilde{\rho}_m(k)dk$ over the density of states $\tilde{\rho}(k)$ per interval $dk$ of the component $k$ of the wave vector parallel to the layers of the structure. Note that the integrals not only include the density of states involving radiation of photons, but also the states involving near-field for $k > k_0$. Generic expressions for $\tilde{\rho}_e(k) \propto dP_e/dk$ and $\tilde{\rho}_m(k) \propto dP_m/dk$ are provided in the Appendix (please see Eqs. (8) and (10)) and were derived following Ref. [23].

In Fig.6 (a) we show the normalized relaxation rates $F_e$ and $F_m$ as functions of distance $h$ from a metal film with thickness $a = 50$ nm. When emitters are placed next to the metal, the emission rate is strongly enhanced, especially for the electric dipole. Such modification of the dipole emission near an interface can be described in terms of the image model [5,9-10]. In our case the dipoles are placed next to metal interface and the frequency range of the emission is close to the conditions of plasmon resonance of the metal, which leads to renormalized Coulomb interaction [24]. At the frequency of plasmon resonance, a source positioned within the near field zone at a distance $h$ from the metal interface induces an image with the amplitude multiplied by a factor

$$\frac{\varepsilon_d - \varepsilon_m}{\varepsilon_d + \varepsilon_m} = \left(-1 + \frac{2i\varepsilon_m'}{\varepsilon_m''}\right) \approx \frac{2i\varepsilon_m'}{\varepsilon_m''}, \tag{1}$$

where $\varepsilon_m'$ is the real part and $\varepsilon_m''$ is the imaginary part of the dielectric permittivity of the metal $\varepsilon_m$, such that $|\varepsilon_m'| \gg \varepsilon_m''$ and $\varepsilon_d$ is the permittivity of the dielectric. This modified image formation can also be understood from the fact that Fresnel coefficient for such an interface in the near-field limit, i.e. at high longitudinal momenta, is $r_p(k_\parallel \to \infty) = 1 - 2i\varepsilon_m'/\varepsilon_m''$ (compare this to the near-field of the super-lens of Ref. [25]). The induced near-fields of the image are produced by plasmonic waves, which destructively interfere far from the dipole and constructively interfere to form the dipole image next to the position of the source dipole.

Formation of the electric dipole image and the dominant contribution of this relaxation channel can be confirmed by the fact that the normalized relaxation rate $F_e$ is directly proportional to $h^{-3}$ for $h < 10$ nm as can be seen from Fig.6 (a). Interaction of the electric dipole with its image results in an increased relaxation rate as well as strong quenching of radiation from emitters positioned near metal films.

Interaction of the magnetic dipole with the near-field created by it is different from that of the electric dipole. The dependence of $F_m$ on $h$ approximately corresponds to $h^{-0.8}$, which first of all means that in the plane geometry there is no near-field image in the form of a magnetic dipole. The near-fields created by a magnetic dipole near plasmonic metal nanostructures is a very interesting problem of optical magnetism, which will be considered elsewhere.

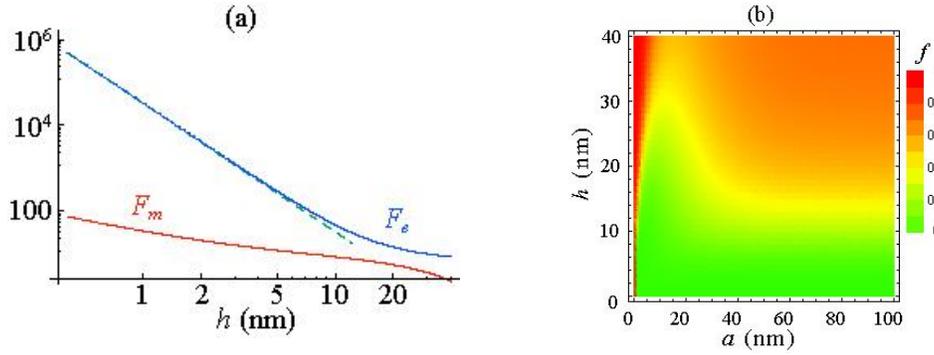

Fig. 6 (a) Normalized total emission rates $F_e$ and $F_m$ of electric and magnetic dipoles placed into a polymer film with $d = 40$ nm next to a gold film with $a = 50$ nm averaged over dipole orientation. (b) Factor $f$ as a function of the metal film thickness and separation of emitter from the metal film $h$ color coded as shown to the right of the graph. The graph is made for $d = 40$ nm, $\beta = 12$ and $\Gamma_{0e}/\Gamma_{0m} = 8$.

The intensity of radiation emitted by the dipoles toward the microscope at an angle $\theta$ to the normal of the structure per solid angle $d\Omega$ is given by

$$\frac{dI_e}{d\Omega} = \frac{P_{uv}}{P_e + P_m + P_{non-em}} \hbar\omega_e \Gamma_{0e} \rho_e(\theta) = \frac{P_{uv}\rho_e(\theta)}{F_e + (\Gamma_{0m}/\Gamma_{0e})F_m + \beta}, \quad (2)$$

$$\frac{dI_m}{d\Omega} = \frac{P_{uv}}{P_e + P_m + P_{non-em}} \hbar\omega_m \Gamma_{0m} \rho_m(\theta) = \frac{P_{uv}\rho_m(\theta)(\Gamma_{0m}/\Gamma_{0e})}{F_e + (\Gamma_{0m}/\Gamma_{0e})F_m + \beta}, \quad (3)$$

where it is assumed that $\omega_m \approx \omega_e$, and $\beta = P_{non-em}/(\hbar\omega_e\Gamma_{0e})$. Here $\rho(\theta)$ is the the local density of states involving emission of a photon in the interval of emission angles $d\theta$ normalized to the density of photons in the vacuum (see Appendix). In our calculations below, we use $\Gamma_{0e}/\Gamma_{0m} \approx 8$ according to the estimations from the experiment (see Fig. 1). We use $\beta$ as the only fitting parameter for our theory.

The numerators in Eqs. (2) and (3) correspond to the far field formation, while the denominators are responsible for the quenching. To illustrate how the quenching is included into our theory we introduce factor $f$, which represents the denominators in Eqs. (2) and (3). The physical meaning of $f$ corresponds to the ratio between the full relaxation rate of emitters on glass to emitters on the metal films. We normalize $f$ by the full relaxation rate on the glass substrate, since this rate is practically independent of $h$

$$f = \frac{(F_e + (\Gamma_{0m}/\Gamma_{0e})F_m + \beta)|_{a \to 0}}{(F_e + (\Gamma_{0m}/\Gamma_{0e})F_m + \beta)}. \qquad (4)$$

The factor $f$ is plotted in Fig. 6 (b). It can be seen that for emitters with $h < 10$ nm $f \approx 0$. This is due to strong quenching, which was described above. Quenching is also stronger for very thin metal films, where it is effective even for emitters separated by $h \approx 30$ nm from the metal. The divide between green and red areas in Fig. 6 (b) defines what we call the "near-field event horizon", beyond which emitters cannot radiate and be detected in the far field.

If an emitter is placed far enough from the metal the quenching is not as strong, which is represented by the factor $f \approx 1$. Those emitters contribute into far field emission and this emission can be explained based on the modified image model. Consider an emitter located next to air-metal interface right at plasmonic resonance. Reflection coefficients for high and low momenta are related as $r_p(k_\parallel \to \infty) = \frac{2r_p(k_\parallel=0)}{1+r_p^2(k_\parallel=0)}$ [26], with reflection coefficient at normal incidence for TM polarization being approximately equal to $r_p(k_\parallel = 0) \approx i$ (the exact equality is in absence of absorption). Thus, the reflection at normal and near normal incidence leads to appearance of phase-shifted image dipoles positioned in metal at distance $h$ from its surface visible in the far-field and observed in the experiment with complex amplitudes

$$\begin{aligned}\boldsymbol{d}_i &= -i\boldsymbol{d}_{0\parallel} + i\boldsymbol{d}_{0\perp},\\ \boldsymbol{m}_i &= i\boldsymbol{m}_{0\parallel} - i\boldsymbol{m}_{0\perp},\end{aligned} \qquad (5)$$

where $\boldsymbol{d}$ and $\boldsymbol{m}$ are correspondingly electric and magnetic dipole moments and subscripts $\parallel$ and $\perp$ correspond to the parallel and perpendicular orientation vs the plane interface.

We explain the properties of the observed emission based on these images. The complex factors in front of the amplitudes lead to a lag in the oscillations of the images with respect to the original dipoles. The radiation emitted by the images travels toward the original dipoles and acquires the corresponding phase. At arrival to the position of the original dipole the emission constructively or destructively interferes with the emission from the original dipole. Since we observe the emission in the direction normal to the interface most of the emission comes from dipoles oriented parallel to the interface and this is where we will place our focus in the discussion.

The amplitude of the waves travelling toward our microscope from an electric dipole next to the metal-dielectric interface is given by

$$1 - i \cdot \exp(2i\varphi_h), \qquad (6)$$

where the phase $\varphi_h = k_0 n_p h$ is related to the propagation from the position of the image to the source. Note that emitters, whose radiation is not quenched, are separated from the metal by distance $h \approx 15 - 40$ nm, while the index of refraction for the polymer $n_p = 1.7$, which makes phase $\varphi_h \approx \pi/15 - \pi/4$. The combination of the quarter-period lag of the image dipole and the phase accumulated during the travel leads to the enhancement of the radiation from the electric dipoles positioned parallel to metal films. This in contrast with the image model based on an ideal reflector, but in agreement with the experimental data.

Now let us turn to the magnetic dipole emission. For a magnetic dipole on top of thick metal films enhancement is observed, while emission is decreased on top of thin films. If one reduces the thickness of the metal film to be on the order of the skin-depth $l_s$ the reflection coefficient is changed to $r_p = i \tanh(a/l_s)$ and the far-field image described above is modified, so that its magnitude becomes reduced. Taking this into account the intensity of the magnetic dipole radiation normal to the structure is modified as

$$|1 + i\tanh(a/l_s)\exp(2i\varphi_h)|^2 \approx 1 + \tanh(a/l_s)^2 - 2\sin(2\varphi_h)\tanh(a/l_s).$$

It can be easily seen that, for example, for $\varphi_h \approx \pi/15$ this function represents enhancement for thick metal films $a \gg l_s$ and reduction of intensity for thin films $a \approx l_s$. It needs to be noted that the reflection characteristics of our actual structure (see Fig. 5) are more complex than the ones we use for the explanations we provide above, first of all, because the emission frequencies of $Eu^{3+}$ transitions are somewhat detuned from the plasmonic resonance. Another factor is the additional reflections from the polymer-air and metal-glass substrate interfaces.

Now having established the groundwork for the theoretical description we turn to the exact situation with which we are presented experimentally. To find the intensities $I_e$ and $I_m$ measured by the microscope, we integrate Eqs. (2) and (3) over the radiative angle from 0 to $\theta_m$ corresponding to the numerical aperture of the microscope $NA = 0.5$. We also average the result over the position $h$ of the emitters within the polymer films. We define the intensity contrast between emitters on metal films and emitters placed directly on the glass substrate as

$$\eta(a) = \frac{I_e(a)}{I_e(a=0)} - 1 \text{ and } \mu(a) = \frac{I_m(a)}{I_m(a=0)} - 1. \tag{7}$$

With this definition a positive value of contrast means that the signal coming from the emitters placed on gold films is stronger than the signal coming from those on the glass substrate. Negative contrast signifies the opposite situation.

The contrast ratios $\eta$ and $\mu$ are shown as functions of the metal film thickness $a$ for gold and silver in Fig. 7. The contrast $\eta$ is positive for gold films thicker than $a \approx 20$ nm and is higher for thicker films, which agrees with the experimental results.

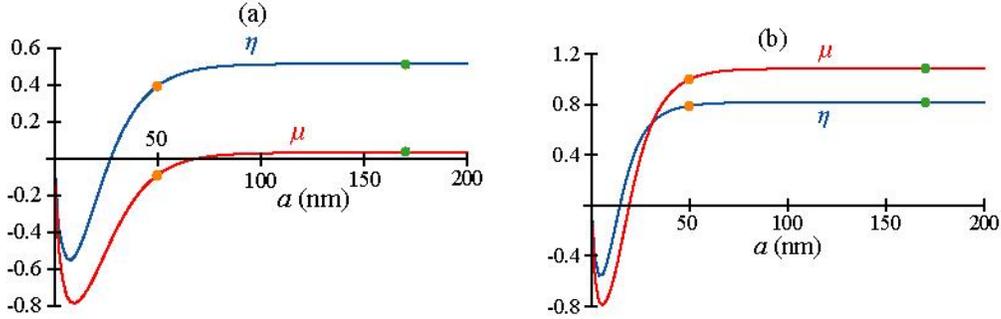

Fig. 7 (a) The contrast ratios $\eta$ and $\mu$ (see Eq. (7)) for electric and magnetic transitions for gold film as a function of film thickness. (b) The same for silver film.

One can also see that the contrast ratio $\mu$ is negative for thin gold films with $a \leq 70$ nm and is positive for thicker films. We have placed orange and green dots in Fig. 7 to highlight the theoretical values corresponding to metal thickness, $a = 50$ nm and $a = 170$ nm, at which the experiments were conducted. It can be easily seen from Fig. 3 (b) and (c) that these contrast ratios correspond nicely to the experimental values. At the same time both $\eta$ and $\mu$ are positive for silver films thicker than 20 nm in accordance with the experiments shown in Fig. 4 (b) and (c). It is through varying the fitting parameter $\beta$ that we establish a curve for the function $\mu$ shown in Fig. 7 (a) that fits the experimental data for gold (Fig. 3 (c)), therefore locking down the value for $\beta = 12$, which seems to be reasonable for our highly luminescent material. Such sensitivity of magnetic dipole radiation to changes in the optical nanoscale environment can serve as yet another proof of the importance of investigations into the field of optical magnetism.

## 4. Conclusions

In conclusion, we have studied the effects of close vicinity to metal on spontaneous emission of electric and magnetic dipole sources through an optical microscope setup. Distinctly different behavior of electric and magnetic dipoles was demonstrated near gold films of a nanoscale thickness. We described the results theoretically based on the dyadic Green's function approach for layered media and proposed an interpretation based on modified image models for the near and far-field.

These results can find applications in probing and mapping of optical field distributions in plasmonic systems by spectroscopic methods.

## 5. Appendix

The electric local density of states $\tilde{\rho}_e(k)$ per interval $dk$ of the component $k$ of the wave vector parallel to the layers of the structure can be found to be

$$\tilde{\rho}_e(k) = \frac{1}{\hbar\omega_e \Gamma_{0e}} \frac{dP_e}{dk}, \qquad (8)$$

where $P_e = ck_0/2 \cdot \text{Im}(\boldsymbol{d}_e^* \hat{G}_e(\boldsymbol{r}_0, \boldsymbol{r}_0)\boldsymbol{d}_e)$, and $\hat{G}$ is the electric dyadic Green's function at the position of the emitter $\boldsymbol{r}_0$. Using the Fourier representation of the Green's function we find

$$\frac{dP_e}{dk} = \frac{ck_0}{2n_p^2} \cdot \text{Re}\left[\frac{k}{k_z}\left(d_{e\parallel}^2 \cdot \frac{k_0^2 n_p^2}{2} \frac{\left(1 + r_{1s} e^{2ik_z(d-h)}\right)\left(1 + r_{2s} e^{2ik_z h}\right)}{1 - r_{1s} r_{2s} e^{2ik_z d}} + d_{e\parallel}^2 \right.\right.$$
$$\left.\left.\cdot \frac{k_z^2}{2} \frac{\left(1 - r_{1p} e^{2ik_z(d-h)}\right)\left(1 - r_{2p} e^{2ik_z h}\right)}{1 - r_{1p} r_{2p} e^{2ik_z d}} + d_{e\perp}^2 \right.\right. \qquad (9)$$
$$\left.\left.\cdot k^2 \frac{\left(1 + r_{1p} e^{2ik_z(d-h)}\right)\left(1 + r_{2p} e^{2ik_z h}\right)}{1 - r_{1p} r_{2p} e^{2ik_z d}}\right)\right].$$

Here $k_z = \sqrt{k_0^2 n_p^2 - k^2}$, $r_{1p}$ and $r_{1s}$ are reflection coefficients from the polymer-air interface for TM and TE polarized radiation, while $r_{2p}$ and $r_{2s}$ are corresponding reflection coefficients for reflection from the metal films.

The magnetic local density of states $\tilde{\rho}_m(k)$ can be found as

$$\tilde{\rho}_m(k) = \frac{1}{\hbar\omega_m \Gamma_{0m}} \frac{dP_m}{dk}, \qquad (10)$$

where $P_m = ck_0 n_p^2/2 \cdot \text{Im}(\boldsymbol{d}_m^* \hat{G}_m(\boldsymbol{r}_0, \boldsymbol{r}_0)\boldsymbol{d}_m)$, and $\hat{G}$ is the magnetic dyadic Green's function. Finally, we find that

$$\frac{dP_m}{dk} = \frac{ck_0}{2} \cdot \text{Re}\left[\frac{k}{k_z}\left(d_{m\parallel}^2 \cdot \frac{k_0^2 n_p^2}{2} \frac{\left(1 + r_{1p} e^{2ik_z(d-h)}\right)\left(1 + r_{2p} e^{2ik_z h}\right)}{1 - r_{1p} r_{2p} e^{2ik_z d}} + d_{m\parallel}^2 \right.\right.$$
$$\left.\left.\cdot \frac{k_z^2}{2} \frac{\left(1 - r_{1s} e^{2ik_z(d-h)}\right)\left(1 - r_{2s} e^{2ik_z h}\right)}{1 - r_{1s} r_{2s} e^{2ik_z d}} + d_{m\perp}^2 \right.\right. \qquad (11)$$
$$\left.\left.\cdot k^2 \frac{\left(1 + r_{1s} e^{2ik_z(d-h)}\right)\left(1 + r_{2s} e^{2ik_z h}\right)}{1 - r_{1s} r_{2s} e^{2ik_z d}}\right)\right].$$

The intensity of radiation emitted into the air by the electric dipole with moment $d_e$ and frequency $\omega = ck_0$ into a solid angle $d\Omega$ at angle $\theta$ to the normal in the far-field zone is

$$\frac{dI_e}{d\Omega} = \frac{c}{4\pi} \overline{|E(\theta)|^2} r^2 = \hbar\omega\Gamma_{0e}\rho_e(\theta). \tag{12}$$

Here $\overline{|E|^2}$ is the electric field at distance $r$ from the sample averaged over the orientation of the dipoles. The spontaneous emission rate in a homogenous polymer medium is $\Gamma_{0e} = \frac{4}{3\hbar n_p^2} k_0^3 |d_e|^2$ and the local density of states involving emission of a photon into the air in the interval of emission angles $d\theta$ normalized to the density of photons in the vacuum for the electric dipole is

$$\rho_e(\theta) = \frac{1}{16\pi} \frac{\cos^2\theta}{(n_p^2 - \sin^2\theta)} \Big(|t_{s+}|^2 + |t_{p+}|^2 \sin^2\theta + |t_{p-}|^2 (n_p^2 - \sin^2\theta)\Big). \tag{13}$$

Similarly, the intensity of radiation by the magnetic dipole reads as

$$\frac{dI_m}{d\Omega} = \frac{c}{4\pi} \overline{|H(\theta)|^2} r^2 = \hbar\omega\Gamma_{0m}\rho_m(\theta). \tag{14}$$

Here the spontaneous emission rate for a magnetic dipole in a homogeneous polymer medium is $\Gamma_{0m} = \frac{4}{3\hbar} k_0^3 |d_m|^2$ and the normalized density of states for the magnetic dipole is

$$\rho_m(\theta) = \frac{1}{16\pi} \frac{\cos^2\theta}{(n_p^2 - \sin^2\theta)} \Big(n_p^4 |t_{p+}|^2 + |t_{s+}|^2 \sin^2\theta + |t_{s-}|^2 (n_p^2 - \sin^2\theta)\Big). \tag{15}$$

The amplitudes of the detected radiation $t_{p,s\pm}$ in Eqs. (12) - (15) are

$$t_{p,s\pm} = \frac{t_{21} \exp(i\varphi_{d-h}) [1 \pm R(a) \exp(2i\varphi_h)]}{1 + r_{12} R(a) \exp(2i\varphi_d)}, \tag{16}$$

$$R(a) = \frac{r_{23} + r_{34} \exp(-\phi_m(a))}{1 + r_{23} r_{34} \exp(-\phi_m(a))}, \tag{17}$$

where phases are given by $\varphi_x = k_0 n_p x$, and $\phi_m(a) = 2k_0 \sqrt{-\varepsilon_m} a \approx 2a/l_s$. The skin-depth is equal to $l_s = (k_0 \text{Re}\sqrt{-\varepsilon_m})^{-1} \approx 25$ nm at optical frequencies. The subscripts in the Fresnel coefficients for p-polarization $r_{ij} = \frac{n_i - n_j}{n_i + n_j}$ and $t_{ij} = \frac{2n_j}{n_i + n_j}$ correspond to the notations given in Fig. 5, while the Airy coefficient $R(a)$ represents reflection from the metal film [27]. The coefficients $t_{p,s\pm}$ contain all the information about the environment in which the emitters are located.

**Acknowledgments**

The work was partially supported by the NSF PREM grant # DMR 1205457, NSF IGERT grant #DGE 0966188, AFOSR grant # FA9550-09-1-0456 and a student research grant from College Office of Undergraduate Research (COUR) at Georgia Southern University.